\documentclass[a4paper,onecolumn,12pt,assc]{IEEEtranASSC}
\usepackage[pdftex]{graphicx}
\def\reff@jnl#1{{\rm#1\/}}
\def\jgr{\reff@jnl{J.~Geophys.~Research}} 
\def\aj{\reff@jnl{Astron.~J.}}          
\def\araa{\reff@jnl{Ann.~Rev.~Astron.~Astrophys}}   
\def\apj{\reff@jnl{Astrophys.~J.}}      
\def\apjl{\reff@jnl{Astrophys.~J.~Lett.}}  
\def\apjs{\reff@jnl{Astrophys.~J.~Suppl.~Ser.}} 
\def\ao{\reff@jnl{Appl.Opt.}}           
\def\aap{\reff@jnl{Astron.~Astrophys.}} 
\def\aapr{\reff@jnl{Astron.~Astrophys.~Rev.}} 
\def\aaps{\reff@jnl{Astron.~Astrophys.~Suppl.}} 
\def\baas{\reff@jnl{Bull.~Am.~Ast.~Soc.}} 
\def\expast{\reff@jnl{Exp.~Astron.}}    
\def\mnras{\reff@jnl{Mon.~Not.~R.~Ast.~Soc.}} 
\def\pasp{\reff@jnl{Pub.~Astron.~Soc.~Pac.}} 
\def\pra{\reff@jnl{Phys.Rev.A}}         
\def\prb{\reff@jnl{Phys.Rev.B}}         
\def\prc{\reff@jnl{Phys.Rev.C}}         
\def\prd{\reff@jnl{Phys.Rev.D}}         
\def\prl{\reff@jnl{Phys.Rev.Lett}}      
\newcommand\nat{\reff@jnl{Nature}}      
\def\procspie{\reff@jnl{Proc.~SPIE}}    
\def\josa{\reff@jnl{J.~Opt.~Soc.~Am.}}   
\def\azh{\reff@jnl{AZh}}                 
\def\apss{\reff@jnl{Ap\&SS}}             

\begin{document}
%
\title{VSTAR Modelling of the Infrared Spectrum of Uranus}


\author{\IEEEauthorblockN{Kimberly Bott\IEEEauthorrefmark{1},
Lucyna Kedziora-Chudczer\IEEEauthorrefmark{1} and Jeremy Bailey\IEEEauthorrefmark{1}}

\IEEEauthorblockA{\IEEEauthorrefmark{1}
School of Physics,
University New South Wales, NSW, Australia, 2052}}

\maketitle

\begin{abstract} 
We modelled the H-band spectrum obtained with the Infrared Imaging
Spectrograph (IRIS2) at the Anglo-Australian Telescope in order to infer the cloud structure and
composition of the  atmosphere of the planet between 0.1 and 11 bar. Such modelling can be used to derive the D/H ratio in the atmosphere of Uranus, which is an important diagnostic of the conditions in early history of the planetary formation.  We describe here our modelling technique 
based on the Versatile Software for Transfer of Atmospheric Radiation (VSTAR) \cite{Bailey2012a}. Since
the infrared spectrum of Uranus is dominated by absorption from methane, the accuracy of the models is
limited largely by the quality of the low temperature, CH$_{4}$ line databases used. Our modelling includes
the latest laboratory line data for methane described in Bailey 2011
\cite{Bailey2011}. The parameters of this model will be applied in the future to model our observations of
GNIRS high resolution spectra of Uranus and derive the D/H ratio in the 1.58$\mu$m window based on
absorption in the CH$_{3}$D band in this region. \end{abstract} 


\section*{Introduction}

The gas giant Uranus, with its 98$^\circ$  spin-axis inclination \cite{2007Seidelmann} has the largest variation in the amount of solar radiation received of any planet as one pole pointing nearly to the sun at a point in the orbit, becomes the pole pointing nearly away from the sun at the opposite point in the orbit. However the lack of an internal heat source combined with the long radiative
time constant is the most likely reason that its atmosphere appears rather devoid of the
band-like features characteristic for Jupiter or Saturn \cite{Sromovsky2005}. The aquamarine colour of the planet is attributed
to methane absorption in visible light. 

After the flyby of Voyager~2 in 1986 \cite{1987Lindal}, detailed imaging and spectroscopy
measurements confirmed that molecular hydrogen and helium (0.152 molar fraction) are the main constituents
of the planet's atmosphere. The next most abundant component is methane with a mixing ratio of about 2\%,
followed by traces of ammonia, water and hydrogen sulphide. The radio occultation measurements suggested
the existence of a thick methane cloud layer in the troposphere at 1.2 bar. Methane freezes out in the
higher levels of the atmosphere and its mixing ratio drops below $5\times 10^{-3}$ at pressures lower than
500mb. A haze of complex hydrocarbons in the stratosphere (at pressures below 20 mbar) was suggested by Spitzer
Telescope 20 $\mu$m observations \cite{Burgdorf2006} and occultation measurements  with the UV spectrometer
on Voyager 2 \cite{Bishop1990}.   Clouds of water, ammonia and hydrogen sulphide are expected to form
in the lower troposphere at pressures above 3 bar. The top of the highest deck of ammonia or hydrogen
sulphide clouds was derived based on the optical spectrum in Irwin 2010 \cite{2010Irwin}.   

Discrete clouds were first imaged by Voyager 2 \cite{Smith1986} and later by Keck and Hubble Space
telescopes \cite{Sromovsky2009} in the near-infrared range. The highest clouds at 200 mbar were visible in the
K-band, while imaging and J and H band spectroscopy showed cloud features at 2 to 5 bar in so-called "methane windows", where the atmosphere is transparent to almost 10 bar. Observations of discrete clouds
allow measurement of the atmospheric circulation and zonal wind speeds at different heights. Seasonal and
longitudinal variations in albedo due to inhomogeneity of aerosols over the disk of Uranus have been
detected and discussed in Kostogryz 2007 \cite{2007Kostogryz}. 

Karkoshka and Tomasko 2009  \cite{Karkoschka2009} analysed a number of spatially resolved spectra of Uranus
obtained with the STIS Hubble spectrograph in a range between 0.3 and 1 $\mu$m and found evidence of
changes in the mixing ratio of methane at different latitudes, with higher mixing ratios at lower latitudes.
This is consistent with the suggestion of a meridional flow with descending gas at high latitudes
\cite{Sromovsky2011}.  Irwin 2010 \cite{Irwin2012} used the latest methane line data (available from Campargue 2012
\cite{Campargue2012}) to model the near-infrared spectra from Gemini North NIFS data to better constrain
methane abundance, the height of the top of the main cloud and the CH$_{3}$D/CH$_{4}$ ratio. Their results
suggest that the thickness and the height of the main cloud varies as a function of latitude. The cloud deck
appears highest at around 45$^{\circ}$ North and South of equator and it thickens towards the equator. 

In this paper we describe our collection and reduction methods in the "Observations" section.  We then describe the Fortran code, VSTAR, used to model a comparison spectrum and describe the considerations that went into setting the model up such as the methane mixing ratio and why it is important, other sources of absorption, scattering sources and the structure of the clouds.  We then discuss the deuterium abundance and its importance before concluding with comments on the abilities of VSTAR, the importance of deuterium measurement and the possibility of better confining the deuterium amounts with new, higher resolution spectra. 


\section*{Observations} On the 31st of July 2010 we acquired long-slit (7.7 arcmin) spectra of Uranus with
the InfraRed Imaging Spectrograph (IRIS2) at the Anglo-Australian Telescope. The observations were taken
with a resolving power of  R$\sim$2400 over the J, H and K bands, which covers the wavelength region between 1.09 to 2.4${\mu}$m.  For the H band, of which this paper is primarily concerned the signal to noise ratio was sufficiently high that it did not affect the quality of the fits.
The 1 arcsec wide slit was oriented north-south on the sky, but the axis of Uranus is at a position angle
of 254$^\circ$ (measured E from N), so our slit crosses mostly equatorial regions of the planet. The width of the slit is
equivalent to 2.2 pixels on the detector. Our observing strategy was to take a sequence of four exposures
(A, B, B, A) where A and B denote frames with the planet offset to two different positions on the slit. Observations were taken in 16 cycles of 1.5s x 41 exposures giving us a total exposure time of 960 seconds.

The spectra were extracted using the standard data reduction package FIGARO \cite{Shortridge1995}. We
carried out flat-field correction of the 2D spectra in order to remove the non-uniform response of the
detector. Wavelength calibration was achieved with a xenon line source. Sky subtraction was performed by
subtracting two corresponding frames with offset spectra (A and B) from each other. We also corrected for
spectral curvature by resampling the frames using the spectra of standard stars taken for each band. In
addition we removed bad pixels due to cosmic rays and known imperfections of the detector by interpolating
between their neighbours. Our sun-like, G-type, standard stars  used for flux calibration and removal of telluric
absorption were observed at close angular proximity to the planet. The absolute radiance factor, I/F, which compares the planet's reflected intensity to its received flux over given wavelengths (as discussed in \cite{Kedziora2011}), is
derived by dividing the extracted spectrum of Uranus by the spectrum of the standard star. 

\begin{figure}[!h]
	\centering
	\includegraphics[width=6.5in]{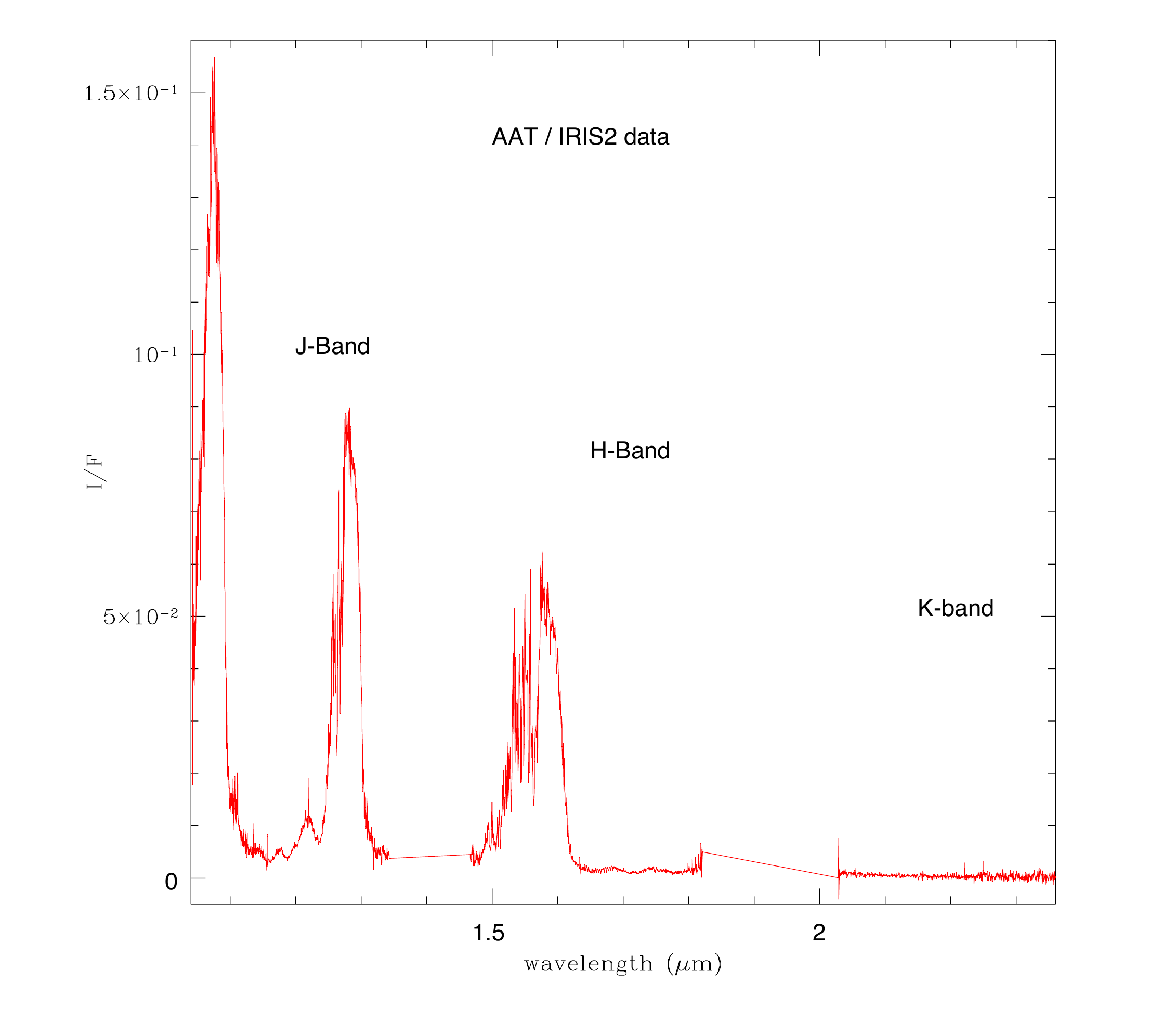}
	
	\caption{The spectrum of Uranus in the near-infrared region observed with the IRIS2 instrument at the AAT on 31/07/10 with a resolution of 2400}
	\label{figure1}
	\end{figure}

Figure~\ref{figure1} shows the resulting spectrum at the J, H and K bands.

\section*{VSTAR model of the H-band spectrum} To obtain the best fit for our observed spectrum we used the
Versatile Software for Transfer of Atmospheric Radiation (VSTAR), which is a modular, line-by-line
atmospheric radiative transfer package described in Bailey and Kedziora-Chudczer 2012 \cite{Bailey2012a}. It has been successfully applied to planets of
the Solar system: Earth \cite{Hough2006}, Venus \cite{Bailey2008} \cite{Bailey2009},  Jupiter
\cite{Kedziora2011}; planetary moons such as Titan \cite{Bailey2011} \cite{Bailey2012} 
\cite{Kedziora2013} and brown dwarfs \cite{Bailey2012a}. The newly extended version of the software also allows modelling the
spectra of transiting extrasolar planets and it can solve for fully polarised radiation transfer.

VSTAR is a Fortran code that calculates the radiative transfer of light in an atmosphere that can have a
multilayer structure, where each layer has a defined chemical composition and scattering particle
properties. The VSTAR modelling code is organised as a Fortran subroutine library from which a specific
model is built by writing a simple program that calls appropriate routines. The MOD package defines a two dimensional grid of parameters for each
atmospheric layer as a function of wavelength or a wave number. The parameters of the atmosphere such as
pressure and temperature for every layer are either known from direct measurements or can be solved for in
a similar way as described in Hubeny 1995 \cite{Hubeny1995}. The LIN package is used to read spectral lines from line databases \cite{Bailey2011} and the
RAY package derives optical depth due to Rayleigh scattering due to molecules. 
The PART module is used next to derive
scattering properties of the aerosols and clouds. In this module, one specifies size distributions and
refractive indices and optical depths for all particles present in different layers of the atmosphere and
the scattering properties are calculated using Lorenz-Mie theory. The major component of the 
VSTAR package is the RT module, which contains a selection of radiative transfer solvers including DISORT
adapted from Stanmes 1988 \cite{Stanmes1988}. VSTAR was recently upgraded to allow a full four Stokes treatment of
the polarized radiative transfer problem \cite{Kedziora2011a}.

The spectrum of Uranus was modelled by dividing the atmosphere into 58 layers ranging from 0 to 136.5 km (or
11285 to 101 mbar) with the temperature-pressure (P-T) profile derived from radio occultations measured
with Voyager 2 \cite{Lindal1987}. For this model we concentrated on the window surrounding 1.6${\mu}$m because the K band for Uranus is low flux and databases for methane lines at the temperature of Uranus are not yet well developed for the J band.

\subsection*{Methane Mixing Ratio Profile and Absorption} 
\begin{figure}[!h]
		\centering
		\includegraphics[width=5.5in]{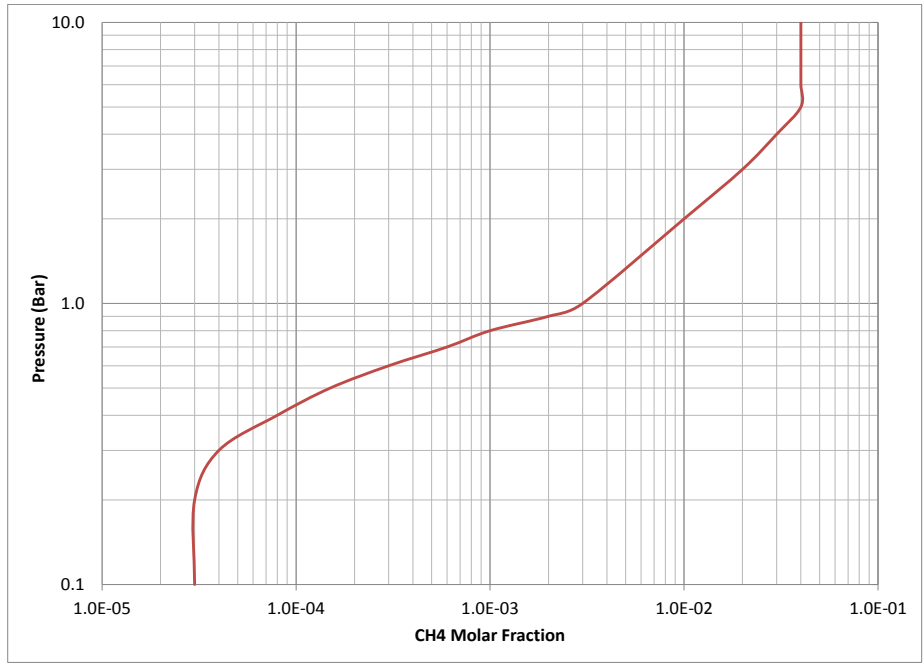}

		\caption{Methane mixing ratio in the atmosphere of Uranus as a function of  pressure.}
		\label{figure2}
		\end{figure}

The distribution of methane over these altitudes
is not uniform and the profile of the methane mixing ratio almost certainly depends on the observed latitudes as
well. The new, refined profile derived from Irwin et al. 2010 \cite{Irwin2010} (Figure~\ref{figure2}) gave the best fit to our data. The CH$_{4}$ mixing ratio at the deepest levels of the atmosphere is 4\%, which
is higher than the 1.6\% estimated from observations at polar latitudes at pressures higher than 1 bar
\cite{Karkoschka2010}. This higher 4\% mixing ratio is consistent with the  P-T profile containing
warmer lower layers of the atmosphere from occultation data \cite{Lindal1987}, as can be seen in the left panel of
Fig. 7 of Irwin 2010 \cite{Irwin2010}. Our slit, positioned at the centre of the planetary disk,
covers a only a limited range of latitudes around the equator. Therefore the single methane profile from Irwin 2010
\cite{Irwin2010} is expected to account for the observed absorption features visible in our data.

Methane is the dominant absorbing species in the atmosphere of Uranus in the near infrared region. The CH$_{4}$
molecule has a complex absorption spectrum, which arises due to four vibrational modes interacting with
each other, forming so-called polyads. Although there has been continuing progress in determination of
the methane spectrum (see \cite{Bailey2012}), the high-order polyads, which form absorption lines in the
near infrared (1 -- 2 $\mu$m) are not well understood. The HITRAN database at these wavelengths is largely
based on empirical measurements at room temperature, but because their
lower-state energies and quantum identifications are missing, it is difficult to model such lines for
different temperatures. 


A recent major breakthrough has been a set of deep laboratory measurements of methane lines at room temperature
and cryogenic temperatures by the Grenoble group (\cite{Wang2011}, \cite{Campargue2010a}) which provide an
excellent line list over the range 1.26 -- 1.71 $\mu$m. These line data have been shown to give
good models for the spectra of Titan (\cite{Bailey2011}, \cite{Bailey2012}, \cite{deBergh2012}). We use a
line list based on this new data, as well as other sources at longer wavelengths, as described in Bailey 2012
\cite{Bailey2012}.


The shape of methane lines, especially in far wings of the strong lines has an effect on the shape of the
absorption windows. In the H-band data we found that the best fit is achieved by using a Lorentzian profile
with a cut-off at 100 cm$^{-1}$. 

\subsection*{Other absorbers} The dominant component of the Uranian atmosphere is molecular hydrogen, which is
not only the source of Rayleigh scattering, but also absorbs infrared radiation due to collisional
interactions in a dense gas that induce electric dipoles that emit or absorb radiation at rotovibrational
frequencies. This collision-induced absorption (CIA) gives rise to a spectrum of overlapping lines, which
give the impression of a smooth absorption trough. Since the atmosphere of Uranus is about 15\% 
helium, one has to consider also collisional interactions between H$_{2}$ and He molecules. We modelled
both in our code based on coefficients derived in Borysow 1991 \cite{Borysow1991}. We note that CIA effects in H-band
spectrum are likely to be rather small, because this is the region of low CIA absorption coefficients
between two prominent bands: the fundamental and its first overtone.  We have also assumed an equilibrium
ortho/para H$_{2}$-ratio in our model.

\subsection*{Scattering}  Two sources of scattering are considered in our model. Rayleigh scattering cross
sections due to molecules are calculated for an
atmosphere composed of H${_2}$ and He. Rayleigh scattering is most efficient at short wavelengths,
so scattering in our infrared region will occur mainly due to clouds and aerosols that form
hazes.

	
	
	\begin{figure}[!h]
	\centering
	\includegraphics[width=6.5in]{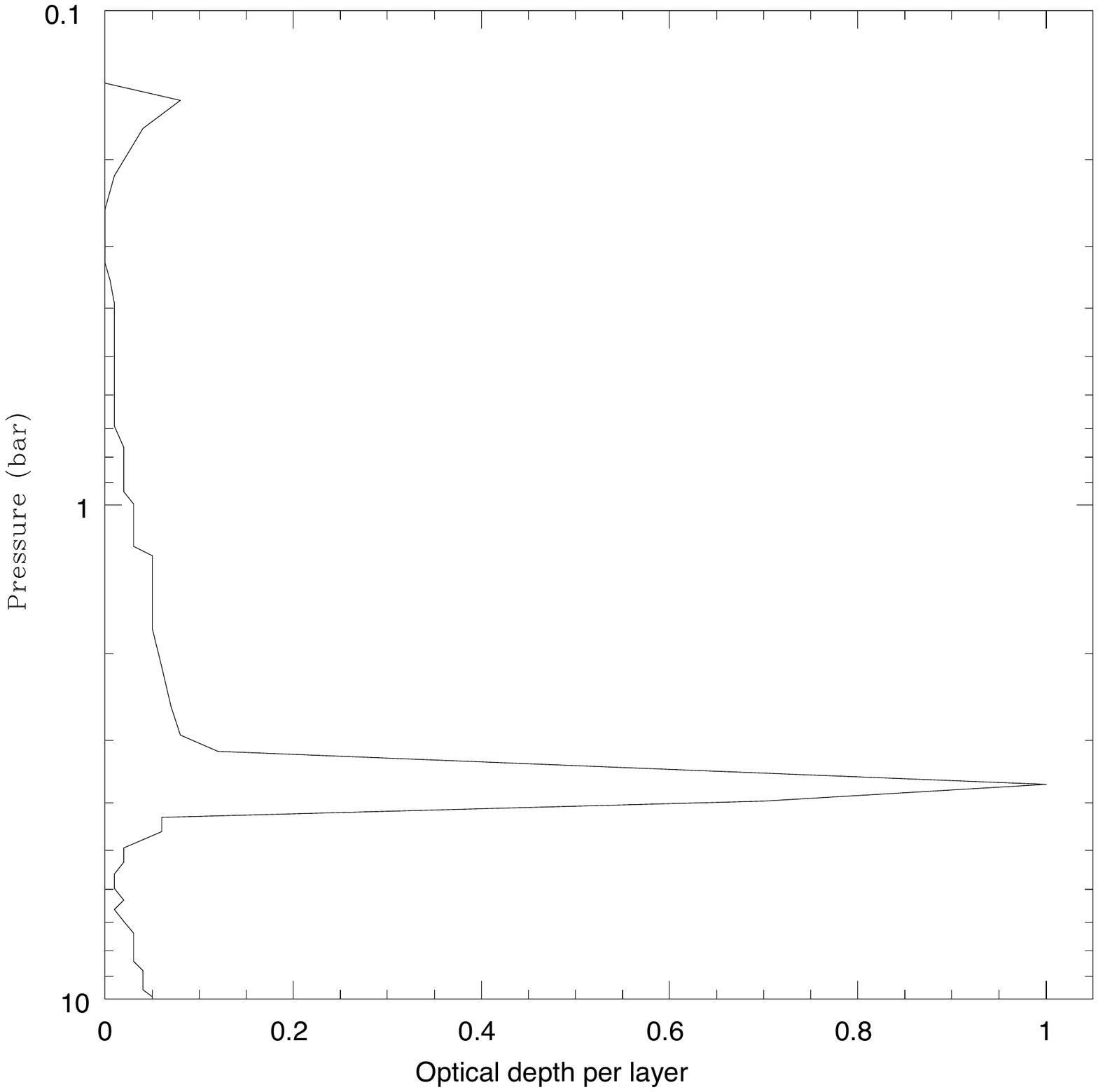}
	
	\caption{Cloud opacity profile used for the model.}
	\label{figure3}
	\end{figure}
	
	Lorenz-Mie theory is used to model scattering from the clouds. We assumed a power-law
	size distribution of particles with effective radius of 2$\mu$m and variance of 0.4$\mu$m with a
	constant  complex refractive index of $n=1.4+0i$ in the considered range of wavelengths. The phase
	function was modelled with a Henyey-Greenstein function. The cloud composition and particle size can
	change by height but we found no major effect on the spectrum by changing  between 1.2 and
	5${\mu}$m sized particles. 

	The cloud optical depth profile presented in Figure~\ref{figure3} is consistent with the latest model
from Irwin 2010 \cite{Irwin2010}, where the 4\% mixing ratio of methane was applied at pressures above 1 bar. This
profile does not have a break in the lower cloud cover in the equatorial regions which was needed to
provide good fit for modelling the data with lower mixing ratios of CH$_{4}$ used by Irwin et al. in the
past \cite{Irwin2007}.  We found that our model spectra required a low opacity haze present at the
pressures from about 0.15 to 0.2 bar. Such a high altitude haze layer was previously necessary for the model
fitting of Uranus used by other groups \cite{Karkoschka2009, Irwin2012}. For simplicity we assumed that the
upper haze was made from particles of the same size as the lower cloud.

Fitting the structure and opacity of the cloud layers was done initially by manual analysis and
manipulation of the opacity profile as a function of pressure. Irwin 2012 \cite{Irwin2012} notes the 
degeneracy between the choice of methane profile and the detailed structure of the cloud decks for
latitudinally averaged spectra.

		

\begin{figure}[!h]
		\centering
		\includegraphics[width=4.5in]{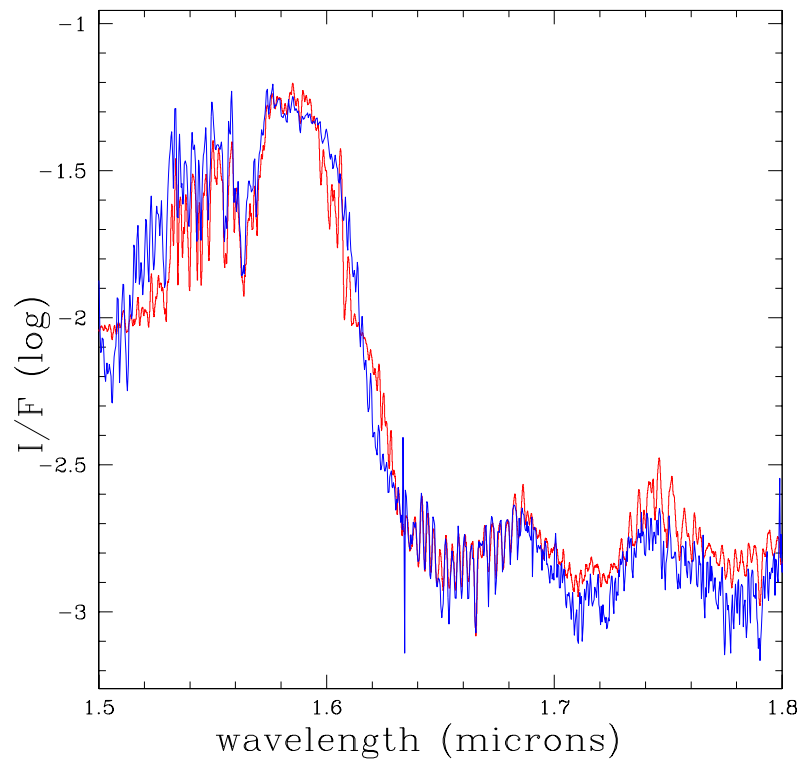}
		\caption{A fit of the near-IR spectrum of Uranus (blue) with VSTAR (red).  The updated methane profile gives better line amplitudes throughout and a much better fit around 1.65 microns.}
		\label{figure4}
		\end{figure}

We compared models with cloud opacity profile adapted from Irwin 2012 \cite{Irwin2012} which
included low altitude haze with the profile adopted from Karkoschka \& Tomasko 2009 \cite{Karkoschka2009} for
a single solid cloud and haze layer at pressures lower than 0.1 bar. Our spectrum was derived from the
narrow range of latitudes around planetary equator, so we applied these cloud profiles to the mixing
ratio profile that includes an enhanced methane abundance at deep levels of atmosphere \cite{Lindal1987}. We found
that the cloud profile similar to one used in Karkoschka 2009 \cite{Karkoschka2009} (Figure~\ref{figure3}) fits our data
better than the profile from Irwin 2012 \cite{Irwin2012}. The corresponding model and the observed spectrum is shown in Figure~\ref{figure4}.

\subsection*{Deuterium Abundances}

The H-band region of the Uranus spectrum can be used to model the ratio of deuterium to hydrogen (D/H).
This ratio is an important diagnostic, helping us to understand the process of  Solar System formation.   The D/H ratio combined with a solar nebula model provides insight into the formation of the giant planets, namely where they formed and through what means their deuterium is likely to have been trapped.  \cite{Hersant2001} \cite{Horner2008}

The new line data from Wang 2011 \cite{Wang2011} provide
identification of the CH$_{3}$D 3$\nu_{2}$ band in the 1.56 $\mu$m window. The most recent modelling of this
band from the Gemini North NIFS data in Irwin 2012 \cite{Irwin2012} provided a D/H ratio in methane of $9.2\times10^{-5}$
as compared with the D/H of the Vienna Standard Mean Ocean Water (V-SMOW) of  $1.56 \times 10^{-4}$.  Deuterium abundances decrease as we move away from the sun in the current Solar System formation model, therefore a vastly different D/H ratio could suggest that Uranus formed far from its current orbital distance.  Uranus' spin-axial tilt, it has been suggested, could be due to collisions with large bodies \cite{Korycansky1990}, so it would not be entirely surprising if it did form at a different distance from where it currently sits and moved die to interactions with large bodies.

The preliminary model in Figure~\ref{figure4} was derived with a D/H of $7.5 \times 10^{-5}$ notably further removed from the Earth abundance than the value of $9.2\times10^{-5}$ from Irwin 2012 {\cite{Irwin2012}. This was used because it provided us with an apparent best fit for the deuterium-sensitive H-band window, however this resolution of spectra is not sufficient to distinguish this value from previously estimated values.  We are
currently analysing high resolution data ($R\sim 18000$) of this region from Gemini North GNIRS from which we should be able to better confine the best fit for the D/H ratio. The model
parameters from the IRIS2 data presented here will be used as starting parameters for the more
detailed and refined model needed to fit the GNIRS spectrum. For icy objects this ratio is determined by the formation temperature and
hence the orbital radius in the Solar System at which ice formed \cite{Horner2007}. Hence, accurate measurements can
provide information on the formation and early history of the ice giant planets, and in particular their
possible orbital migration \cite{Kedziora2013}.
Our GNIRS data should enable improved determination of the D/H ratio due to its higher spectral resolution compared to the AAT IRIS2 data.  The GNIRS observations were also obtained in good seeing conditions, and so have better spatial resolution, which will allow to apply more targeted model parameters for different ranges of latitudes.


\section*{Discussion and Conclusions}

Our results show that a much improved model of the spectrum of Uranus in the H-band region can be achieved
using the new methane spectral line data now available. Because this line data is not available at wavelengths
shorter than 1.26 $\mu$m, it is not currently possible to model the J-band data using these line-by-line
techniques. 

This type of model will be a valuable starting point for modelling the much higher resolution
Gemini GNIRS data on Uranus that we have taken recently. These higher resolution spectra should enable much more
accurate determinations of the D/H ratio and will also allow searching for other trace species that might
be present in the atmosphere.

Uranus now joins a growing list of Solar System atmospheres that have been successfully modelled with VSTAR
and shows the versatility of the methods employed by the software. VSTAR allows the modelling of exoplanetary spectra, such as those that are now
being obtained from transits and secondary eclipses. As has been shown in the modelling of
the Solar System giant planets, the quality of the spectral line data is important in achieving successful
results. While the methane line data are now good enough for accurate modelling of low temperature objects like Uranus, the line lists are still not sufficiently
complete for high temperature objects such as hot Jupiters, where hot bands of methane become important.  The design of VSTAR allows the easy inclusion of updated line lists
as they become available, meaning that once these lists are developed, VSTAR will be ready to move on to exotic exoplanets.



We've produced a good fit to the observed spectrum with two cloud layers, the upper one of which being a thin haze-like layer, and an adopted mixing ratio.  Further analysis is needed for the deuterium abundance to be well determined and any firm conclusions drawn about it.  GNIRS observations already acquired will be analysed to pin down the deuterium abundance, although the IRIS2 data we've considered here has given us a firmer understanding of the structure and composition of the atmosphere, allowing us to more confidently fine tune the deuterium abundance.  The
implications of these findings however are of interest as the deuterium abundance can give us clues about
where a planet has formed and under what conditions. \cite{Bailey2011} \cite{Irwin2012} \cite{Kedziora2011} \cite{Kedziora2013} \cite{Horner2007}




\section*{Acknowledgments}
We acknowledge the support of the Australian Astronomical Observatory in scheduling the observations presented here in the IRIS2 observing programme.




\begin{thebibliography}{1}

\normalsize{}

\bibitem{Bailey2012a}
Bailey J. and Kedziora-Chudczer, L.,
``Modelling the spectra of planets, brown dwarfs and stars using VSTAR'',
\emph{MNRAS}, Vol. 419, 2012, 1913-1929

\bibitem{Bailey2011}
Bailey, J.,  Ahlsved, L. and Meadows, V. S.,
``The near-IR spectrum of Titan modeled with an improved methane line list'',
 \emph{Icarus}, Vol. 213, 2011, 218 -232

\bibitem{2007Seidelmann}
Seidelmann, K., Archinal, B.A., A'Hearn, M.F, et al.
``Report of the IAU/IAG Working Group on cartographic coordinates and rotational elements: 2006"
\emph{Celestial Mechanics and Dynamical Astronomy}, Vol. 98, Issue 3, 2007, 155- 180.

\bibitem{Sromovsky2005}
Sromovsky, L.A. and Fry, P.M.
"Dynamics of cloud features on Uranus"
\emph{Icarus}, Vol. 179, Issue 2, 2005, 459- 484.

 \bibitem{1987Lindal}
Lindal, G.F., Lyons, J.R., Sweetnam, D.N., Eshleman, V.R., Hinson, D.P.,
``The atmosphere of Uranus - Results of radio occultation measurements with Voyager 2'', \emph{Journal of Geophysical Research}, Vol. 92, 1987, pp. 14987-15001.

\bibitem{Burgdorf2006}
Burgdorf, M., Orton, G., van Cleve, J., Meadows, V., \& Houck, J.\ 2006, 
``Detection of new hydrocarbons in Uranus' atmosphere by infrared spectroscopy'',
\emph{Icarus}, Vol.184, 2006, 634-637

\bibitem{Bishop1990}
Bishop, J., Atreya, S.~K., Herbert, F., \& Romani, P.\ 1990, 
``Reanalysis of Voyager 2 UVS occultations at Uranus - Hydrocarbon mixing ratios in the equatorial stratosphere'',
\emph{Icarus}, Vol. 88, 1990, 448-464 

 \bibitem{2010Irwin}
Irwin, P.G.J., Teanby, N.A., Davis, G.R., "Revised vertical cloud structure of Uranus from UKIRT/UIST observations and changes seen during Uranus' Northern Spring Equinox from 2006 to 2008: Application of new methane absorption data and comparison with Neptune" 
\emph{Icarus}, Vol. 208, 2010, pp. 913-926.

\bibitem{Smith1986}
Smith, B. A., Soderblom, L. A., Beebe, R., Bliss, D., Brown, R. H., Collins, S. A., Boyce, J. M., Briggs, G. A., Brahic, A., Cuzzi, J. N. \& Morrison, D.
``Voyager 2 in the Uranian System: Imaging Science Results''
 \emph{Science}, Vol. 233, 1986, 43 

\bibitem{Sromovsky2009}
Sromovsky, L.~A., Fry, P.~M., Hammel, H.~B., et al.
``Uranus at equinox: Cloud morphology and dynamics''
\emph{Icarus}, Vol. 203, 2009, 265-286 

\bibitem{2007Kostogryz}
Kostogryz, N.~M. 
``Study of the Reasons for the Geometric Albedo Variations of Uranus''
\emph{Proceedings of the 14th Young Scientists Conference on Astronomy and Space Physics}, 44-48

\bibitem{Karkoschka2009}
Karkoschka, E., Tomasko, M. 
``The haze and methane distributions on Uranus from HST-STIS spectroscopy.'' 
\emph{Icarus} Vol. 202, 2009, 287�309.

\bibitem{Sromovsky2011}
Sromovsky, L. A., Fry, P. M. and J. H. Kim
``High-latitude depletion of methane on Uranus''
\emph{EPSC-DPS Joint Meeting 2011}, EPSC Abstracts Vol. 6, 2011 

\bibitem{Irwin2012}
Irwin, P.~G.~J., de Bergh, C., Courtin, R., et al.
``The application of new methane line absorption data to Gemini-N/NIFS and KPNO/FTS observations of Uranus� near-infrared spectrum''
\emph{Icarus}, Vol. 220, 2012, 369--382

\bibitem{Campargue2012}
Campargue, A.,  Wang, L., Mondelain, D., Kassi, S. et al.,
``An empirical line list for methane in the 1.26-1.71 $\mu$m region for planetary investigations (T = 80-300 K). Application to Titan'',
\emph{Icarus}, Vol. 219, 2012, 110-128

\bibitem{Shortridge1995}
Shortridge K., Meatheringham S. J., Carter B. D. \& Ashley M. C. B.
``Making the FIGARO Data Reduction System Portable''
\emph{ PASA}, Vol.12, 1995, 244

\bibitem{Kedziora2011}
Kedziora-Chudczer, L. and Bailey, J.
``Modelling the near-IR spectra of Jupiter using line-by-line methods''
\emph{MNRAS}, vol.414, 2011,1483-1492

\bibitem{Hough2006}
Hough, J.H., et al. 
``PlanetPol: A Very High Sensitivity Polarimeter''
\emph{Mon. Not. R. Astr. Soc.}, Vol. 118, 2006, 1302-1318.

\bibitem{Bailey2008}
Bailey, J., Meadows, V.S., Chamberlain, S., Crisp, D., 
``The temperature of the Venus mesosphere from O2(a$^1\Delta_g$) airglow observations,''
 \emph{Icarus} 197, 247-259.

\bibitem{Bailey2009}
Bailey, J.
``A comparison of water vapor line parameters for modeling the Venus deep 
atmosphere. "
\emph{Icarus} Vol. 201, 2009, 444-453.

\bibitem{Bailey2012}
Bailey, J.,
``Methane and Deuterium in Titan's Atmosphere'',
\emph{Proceedings of the 11th Australian Space Science Conference}, 2012, Vol. 55, 64 - 76

\bibitem{Kedziora2013}
Kedziora-Chudczer, L., Bailey, J. \& Horner, J.
``Observations of the D/H ratio in Methane in the atmosphere of Saturn's moon, Titan - where did the Saturnian system form?''
\emph{Proceedings of the 12th Australian Space Science Conference}, Submitted

\bibitem{Hubeny1995}
Hubeny, I., \& Lanz, T. ,
``Non-LTE line-blanketed model atmospheres of hot stars. 1: Hybrid complete linearization/accelerated lambda iteration method'', 
\emph{ApJ}, Vol. 439, 1995, 875

\bibitem{Stanmes1988}
Stanmes, K., Tsay, C. S., Wiscombe, W. \& Jayaweera, K.
``A numerically stable algorithm for discrete-ordinate-method radiative transfer in multiple scattering and emitting layers media''
\emph{Appl. Opt},  Vol. 27.1988,  Issue 12, 2502.

\bibitem{Kedziora2011a}
Kedziora-Chudczer, L., Bailey, J., "Application of the VSTAR Code for Modeling of Spectra and Polarization Curves of Hot Jupiters",
 \emph{AAS/Division for Extreme Solar Systems Abstracts}, Vol. 2, 2011, p. 4012.

\bibitem{Lindal1987}
Lindal, G. F., Lyons, J. R., Sweetnam, D. N., Eshleman, V. R., Hinson, D. P.
``The atomsphere of Uranus -- Results of radio occultation measurements with Voyager 2.''
\emph{J. Geophys. Res.} Vol.92 (11), 1087, 14987-15001

\bibitem{Irwin2010}
Irwin, P.G J., Teanby, N. A. and Davis, G. R.
``Revised vertical cloud structure of Uranus from UKIRT/UIST observations and changes seen during Uranus' Northern Spring Equinox from 2006 to 2008: Application of new methane absorption data and comparison with Neptune''
\emph{Icarus}, Vol. 208, 2010, 913-926

\bibitem{Karkoschka2010}
Karkoschka, E., Tomasko, M. 
``Methane absorption coefficients for the jovian planets from laboratory, Huygens, and HST data'' 
\emph{Icarus} Vol. 205, 2010, 674-694.

\bibitem{Wang2011}
Wang, L., Kassi, S., Liu, A.W., Hu, S.M., Campargue, A.,
``The 1.58 $\mu$m transparency window of methane (6165-6750 cm-1): Empirical line list and temperature dependence between 80 and 296 K''
\emph{J. Quant. Spectrosc. Radiat. Trans.}, Vol. 112, 2011, 937-951.

\bibitem{Campargue2010a}
Campargue, A., Wang, L., Kassi, S, Masat, M.,, Votava, O.,
``Empirical line parameters of methane in the 1.63-1.48 $\mu$m transparency window by high sensitivity Cavity Ring Down Spectroscopy''
\emph{Chemical Physics}, vol. 373, 2010, 2034-210.

\bibitem{deBergh2012}
De Bergh, C. et al.,
``Application of a new set of methane line parameters to the modeling of Titan's spectrum in the 1.58
$\mu$m window''. \emph{Plan. Space Sci}, Vol. 61, 2012, pp 85-98.

\bibitem{Borysow1991}
Borysow, A. . 
``Modeling of collision-induced infrared absorption spectra of H2- H2 pairs in the fundamental band at temperatures from 20 to 300 K.
\emph{Icarus}, Vol. 92, 1991, 273- 279.

\bibitem{Irwin2007}
Irwin, P.G J., Teanby, N. A. and Davis, G. R.
``Latitudinal variations in Uranus' vertical cloud structure from UKIRT UIST observations''
\emph{The Astrophysical Journal}, Vol. 665,2007,  L71-L74



\bibitem{Hersant2001}
Hersant, F., Gautier, D., and Hure, J-M.
"A two-dimensional model for the primordial nebula constrained by D/H measurements in the solar system: Implications for the formation of giant planets."
\emph{Astrophysical Journal}, Vol. 554, 2001, 391-407.

\bibitem{Horner2008}
Horner, J., Mousis, O., Aliber, Y., Lunine, J., Blanc, M.
"Constraints from deuterium on the formation of icy bodies in the Jovian system and beyond."
\emph{Planetary and Space Science}, Vol. 56, Issue 12., 2008, 1585-1595.

\bibitem{Korycansky1990}
Korycansky, D.G., Bodenheimer, P., Cassen, P., Pollack, J.B.
"One-dimentional calculations of a large impact on Uranus."
\emph{Icarus}, Vol. 84, Issue 2, 1990, 528-541.

\bibitem{Horner2007}
Horner, J., Mousis, O. and Hersant, F.
"Constraints on the Formation Regions of Comets from their D:H ratios"
\emph{Earth, Moon and Planets}, Vol. 100, Issue 1-2, 2007, pp. 43 - 56

\end{thebibliography}
%

\end{document}